\def\<{\langle}
\def\>{\rangle}
\def\etal{{\it et al. }}
\def\kev{\rm keV}
\documentstyle[12pt,aasms4]{article}
\singlespace
\begin{document}
\title{ The Implications of Direct Red-Shift Measurement of 
$\gamma$-ray Bursts.}
\author{ Ehud Cohen and Tsvi Piran}
\affil{Racah Institute of Physics, The Hebrew University, 
Jerusalem, Israel 91904}

\abstract{  
The recent discoveries of X-ray and optical counterparts for GRBs, and
a possible discovery of a host galaxy, implies that a direct
measurement of the red-shift of some GRBs host galaxies is eminent.
We discuss the implications of such measurements. They could
enable us to determine the GRBs luminosity distribution, the variation
of the rate of GRBs with cosmic time, and even, under favorable circumstances,
to estimate $\Omega$. 
Using GRB970508 alone, and assuming standard candles, we constrain the
intrinsic GRB evolution to $\rho(z)=(1+z)^{-0.5\pm0.7}$.

\section{Introduction}

The recent observations of the Italian-Dutch Beppo/SAX satellite of
$\gamma$-ray bursts, (\cite{beppo}),with error boxes of a few
arc-minutes across, enabled a follow-up by optical and radio
observations and the discovery of x-ray, optical and radio
counterparts to GRBs. The optical observations provided, for the first
time, independent estimates of the distances to GRB sources, using
absorption lines or association with host galaxies, (\cite{redshift}
), and demonstrated beyond doubt the cosmological origin of GRBs.
It is highly possible that in the months to
come several GRBs would have independent red-shifts estimates. We may
use these to obtain estimates of the luminosity diversity and the intrinsic
evolution of the GRB rate with cosmic time. Under favorable conditions
we might even be able to use GRBs as cosmic probes for estimating the
cosmological parameters.

Given a group of GRBs with measured fluxes and red-shifts we should
first calculate the luminosity distribution of the sources . Then
using this luminosity function we should estimate the theoretical
peak-flux distribution in cosmological models. Comparison of this
distribution with the observed one would yield a direct estimate of
the cosmic evolution of GRBs - that is the variability of the GRB rate
per unit comoving volume per unit comoving time. Depending on the
width of the luminosity function, the null assumption of no cosmic
evolution could be proven or ruled out with a few dozen bursts.
Cosmological parameters like the closure parameter, $\Omega$, and the
cosmological constant, $\Lambda$, influence only weakly the peak-flux
distribution (\cite{CP}) hence this analysis can be done safely
assuming $\Omega=1$ and $\Lambda=0$. However, if the luminosity
function is narrow enough we might use the peak-flux red-shift
relation to obtain a direct measure of $\Omega$.

\section{ The  Luminosity Function.} 
                                   
The association of a host galaxy with a GRB provides us with the
red-shift of the host, which is the bursts' red-shift
assuming that the burst is close to the host.
Additionally we have the usual peak photon flux parameter, $p (ph/cm^2/sec)$,
which is transferred
later to apparent luminosity, $l (erg/cm^2/sec)$, using bursts' spectra.
The peak photon flux
vs. red-shift and source luminosity  is:
\begin{equation}
\label{p_of_N}
p={ N\left[{ \nu_1 (1+z) ... \nu_2 (1+z)}\right] \over 4 \pi (1+z)  (2c/H_0)^2 
( 1 - {1\over \sqrt{1+z}})^2 },
\end{equation}
where $\nu_1=50\kev, \nu_2=300\kev$ are the detector boundaries, $z$ is the
burst red-shift, $c/H_0$ is  Hubble distance, and $N[\nu_1,\nu_2]$ is 
the number of
photons emitted in the range $[\nu_1,\nu_2]$. 
We have used $\Omega=1$ and $\Lambda=0$ in eq. \ref{p_of_N}.
The effect of $\Omega$ and $\Lambda$ on the luminosity
is not large, and we will  discuss it  later.
The luminosity dependence on Hubble constant is a 
scale factor of  $h_{75}^2 = ({ H_0 / (75 {Km/sec \over Mpc})})^2 $.

When comparing bursts from different red-shifts one must recall that
the observed peak-flux is in the range $50-300$\kev. This corresponds
to different energy ranges at the sources.  In order to discuss a single
luminosity that classifies the burst, we will
consider $ L \equiv \int_{50\kev}^{300\kev} L_\nu \nu d\nu$ at the
source.  To convert from the observed peak flux to the intrinsic
luminosity we assume that the source spectral form is a power-law
$L_\nu = L (\nu/50\kev)^{-\alpha} (2-\alpha)/( 6^{2-\alpha}-1)/(50\kev)^2$
in the energy range 
$50\kev<\nu<300\kev(1+z_{max})$, so that wherever the source is,
the detector sees a power-law spectra.  We will use $\alpha=1.5$ for
all bursts. 
This
value is probably a good typical estimate ( \cite{band} ),
even though the spectra is not the same for all bursts . If
necessary, one can use the measured spectrum of each burst to estimate
its intrinsic luminosity at the $50-300$\kev band, however at this
stage this simple estimate is sufficient.  
Alternatively one can view the variability in the spectral index as an
additional  random
variable that simply widens the luminosity function.
Using this spectral shape we obtain:
\begin{equation}
\label{l_of_z}
l={ L (1+z)^{-\alpha} \over 4 \pi  (2c/H_0)^2 
( 1 - {1\over \sqrt{1+z}})^2 } =
{ L (1+z)^{-\alpha}h_{75}^2 \over 
 7.7*10^{57} ( 1 - {1\over \sqrt{1+z}})^2  } 
\end{equation}


We would be able to obtain
a direct estimate of the GRB luminosity function in the observed band
when as few as a dozen bursts will be observed.  Using eq. \ref{l_of_z}
we find the luminosity for each burst.  Using those luminosities
we can
estimate the luminosity distribution function using
maximum-likelihood or any other statistical method. 
Recall that
current data, and in particular the peak-flux statistics of GRBs
does not constrain the luminosity distribution of GRBs (
\cite{CP}, \cite{loredo} ).
Clearly, if the sub-group of GRBs with optical counterparts is biased,
this distribution estimate will be biased.

\section{Cosmological  GRB Evolutions } 

One of the interesting features that might distinguish between
different cosmological GRB models is the rate that GRBs occur per unit
time per unit comoving volume: $\rho(z)$.  These new measurements
could yield a direct estimate of this distribution.  Once the GRB
luminosity distribution is known we can proceed and compare the
theoretical peak-flux statistics (using the observed luminosity
distribution) with the observed one.
This distribution depends strongly
on the intrinsic evolution of GRBs, that is on variation of $\rho(z)$.
Following (\cite{CP}) we characterize this dependence as
$\rho(z)=(1+z)^{-\beta}$. Comparison of the theoretical and observed
distribution would limit $\beta$. The cosmological
parameters $\Omega$ and $\Lambda$ influence this distribution
rather weakly (\cite{CP}) and have no substantial effect in estimating $\beta$.
Recall that prior to the independent knowledge of 
GRBs luminosity, one could not distinguish between 
cosmological effects and intrinsic evolution
using count statistics.

In fact this comparison can be done even with the current data and
assuming a narrow luminosity distribution (standard candles). 
We can	use GRB970508 to constrain the evolution.  Using the
peak flux = $1.6*10^{-7}erg/sec/cm^2$, (\cite{flux}), the red-shift 
of the
absorption lines $z=0.835$ (\cite{redshift}) which sets 
a lower limit $z> 0.835$ for the burst,
and the 
absence of prominent Lyman-alpha forest in the spectrum which 
compose an upper limit $z<2.1$, we obtain $\beta=-0.1\pm1.3$ in 99\% confidence
level.
Assuming that the absorption
line of GRB970508 correspond to its own red-shift we estimate 
$\beta=0.5\pm0.7$ with this confidence level , see fig 1.
The simplest hypothesis of no evolution $\beta =0$
is consistent with the observations. 
A milder assumption of Gaussian luminosity
distribution with $\sigma_L = L_{obs}/2$, instead of standard candles, 
gives a lower limit 
$\beta > -0.7$ and no upper limit.

\section{Estimates of Cosmological Parameters.}

Despite numerous attempts to estimate the cosmological parameters,
there are still large uncertainties. One may wonder whether GRBs would
provide a meaningful independent estimate of these parameters.  
Using GRBs count statistics
alone, $\Omega$ could not be estimated from the current data (\cite{CP}). 
However,
given a cosmological distribution of sources with measured red-shifts,
we can try to estimate  the cosmological closure parameter, $\Omega$,
in a similar manner to the attempts to estimate $\Omega$ from type I 
supernovae by \cite{omega}.

The observed peak-flux depends on $\Omega$ as:
\begin{equation}
\label{l_z_omega}
l = l(L,\Omega) = L { (H_0/c)^2 \Omega^4 (1+z)^{2-\alpha}  \over 
 64 \pi ({z\Omega/2 + ( \Omega/2-1) ( \sqrt{\Omega z +1 } -1) })^2 }
\end{equation} 
Using the known parameters of each burst (peak-flux and red-shift) we
obtain for each burst a function $L_i=L_i(\Omega)$.  In the previous
sections we have assumed $\Omega=1$ and obtained
$L_i(\Omega=1)$ ).  For standard candles all $L_i$ must be equal.
Given two sources we have two functions $L=L_{1,2}(\Omega)$, with two
variable and we should be able to determine $\Omega$.

A luminosity distribution will induce an uncertainly in this estimate
that can be approximated by:
\begin{equation}
\sigma_{\Omega}(z) \approx  \left. 
{ d\Omega \over dL } \right|_{\Omega=1,z} \sigma_L = 1/4 {
(\sqrt{1+z}-1) \sqrt{1+z} \over { z/2 - 3/2 \sqrt{1+z} + 3/2 +
z/(4\sqrt{1+z}) }} {\sigma_L \over L}
\end{equation}
For $z=1.5$ we obtain $\sigma_{\Omega} \approx 2 {\sigma_L / L} $. Thus
assuming ${\sigma_L / L}=1$, we need 100 bursts with a measured $z$ to
estimate $\Omega$ with an accuracy of $\sigma_{\Omega} = 0.2$.  Such a
goal could be achieved within several years.  At present it is not
known whether the GRB luminosity distribution is narrow enough and satisfies 
this condition. However, as we have shown at section 2, the width
of the luminosity distribution will be known soon. 

\section{ Future Detectors} 

In view of the promising avenues that these observations have opened
it is worthwhile to examine what will be the effect of future more
sensitive detectors on these estimates. 
Cohen and Piran (1995) have
estimates that BATSE is sensitive to bursts up to $z_{max}=2$. A more
sensitive detector ( by a factor of 10 ) will detect bursts up to
$z=6.9$.
Assuming that $\rho(z)$ is constant up to high value of $z$ we find
that the number of observed bursts will increase only by a factor of
2.1.  
The results are slightly better if the current $z_{max}$ is smaller.
For example a ten fold more sensitive detector will measure 
2.6 times more bursts than BATSE if $z_{max}$=1.5 and 3.5 times more bursts
if $z_{max}=1$.

At first sight these results might look discouraging.  However, the
rate of GRBs at high red-shift is unknown and could be critical in
determining the nature of GRBs. At present it is not known if there
are bursts which originate from high $z$.  Most models that are based
on compact objects cannot produce sources at very early time.  On the
other hand these model, and in particular the neutron star merger
model predict a high rate of GRBs that will follow with a time lag of
up to $~10^9$ years any extended star formation activity.  This is
approximately the time it will take the stars in a binary system to
finish their life cycle, become NS, loose angular momentum through
gravitational radiation, and merge. Nuclear abundances  measurement
indicates that heavy elements that are produced in supernovae began to
be produced earlier than heavy elements produced in neutron star mergers.
It will be intriguing to see whether GRB rates would follow the trend of
supernovae or neutron star mergers.


Furthermore, , one has to recall that the relevant question for our
purposes is not how many bursts are observed but how many bursts are
observed with measured red-shifts. Currently the rate of detection of
bursts with counter-parts is about one per month and from those
detected until now only one has a measured red-shift. Here there is an
enormous potential for improvements. For example systematic
measurements of the red-shift of all bursts observed by BATSE
($\approx 300$ per year) would yield an independent estimate of
$\Omega$, with $\sigma_\Omega=0.1$, even if the luminosity function is
wide, (${\sigma_L / L}=0.9$), within one year.

\eject

\section{Discussion and Conclusions}

As expected, the direct red-shift measure of GRB970508 agrees well
with estimates made previously using peak-flux count statistics
(\cite{fenimore}, \cite{loredo}, \cite{CP}).  It is remarkable what
could be done with even several additional red-shifts. An estimate of
the bursts' luminosity distribution can be obtained even with a few
bursts.  This luminosity function combined with the observed peak-flux
distribution   would provide us immediately with an estimate of the
cosmological evolution of the rate of GRBs.

It is generally accepted that Fireball is inevitable in any
cosmological model. Within this model the observed $\gamma$-rays are
produced during the conversion of a relativistic energy flow to
radiation. However, the source itself that produces the flow remains
unseen (\cite{tsvi}). The limits on cosmological evolution could shed
light on the GRB mystery, by distinguishing between different
cosmological models.  For example the expected rate of merging neutron
stars depends on the red-shift in a drastically different way than the
expected rate of events related to AGNs.

The implications of these red-shift measurements could reach even
further.  Detection of a significant group of GRBs with red-shifts
could enable us to utilize GRBs to study cosmology.  Hundred bursts
with associated red-shifts will enable us to estimate the cosmological
parameter $\Omega$ even if the GRB luminosity distribution is relatively
wide.  This goal is not practical with the current detection
rate. However, if the luminosity distribution is narrow, or if a novel
detection technique could be found which will yield a significantly
higher rate of counter-part detection, we would be able to measure
$\Omega$ using this method within several years.

\section{Acknowledgments}
This research was supported by a US-Israel BSF grant 
and by a NASA grant NAG5-1904.

\eject

\begin{figure}
\caption{ 
The likelihood function (levels 31.6\%, 10\%, 3.16\%, 1\%, etc. of the maximum)
in $(\beta, L)$ plane for standard candles., $\alpha=1.5$, $\Omega=1$,
and evolution given by $\rho(z)=(1+z)^{-\beta}$. Superimposed on this
map is the luminosity of GRB970508, with thick line where the likelihood
function $>1\%$. We have used $h_{75}=1.$
}
\end{figure}

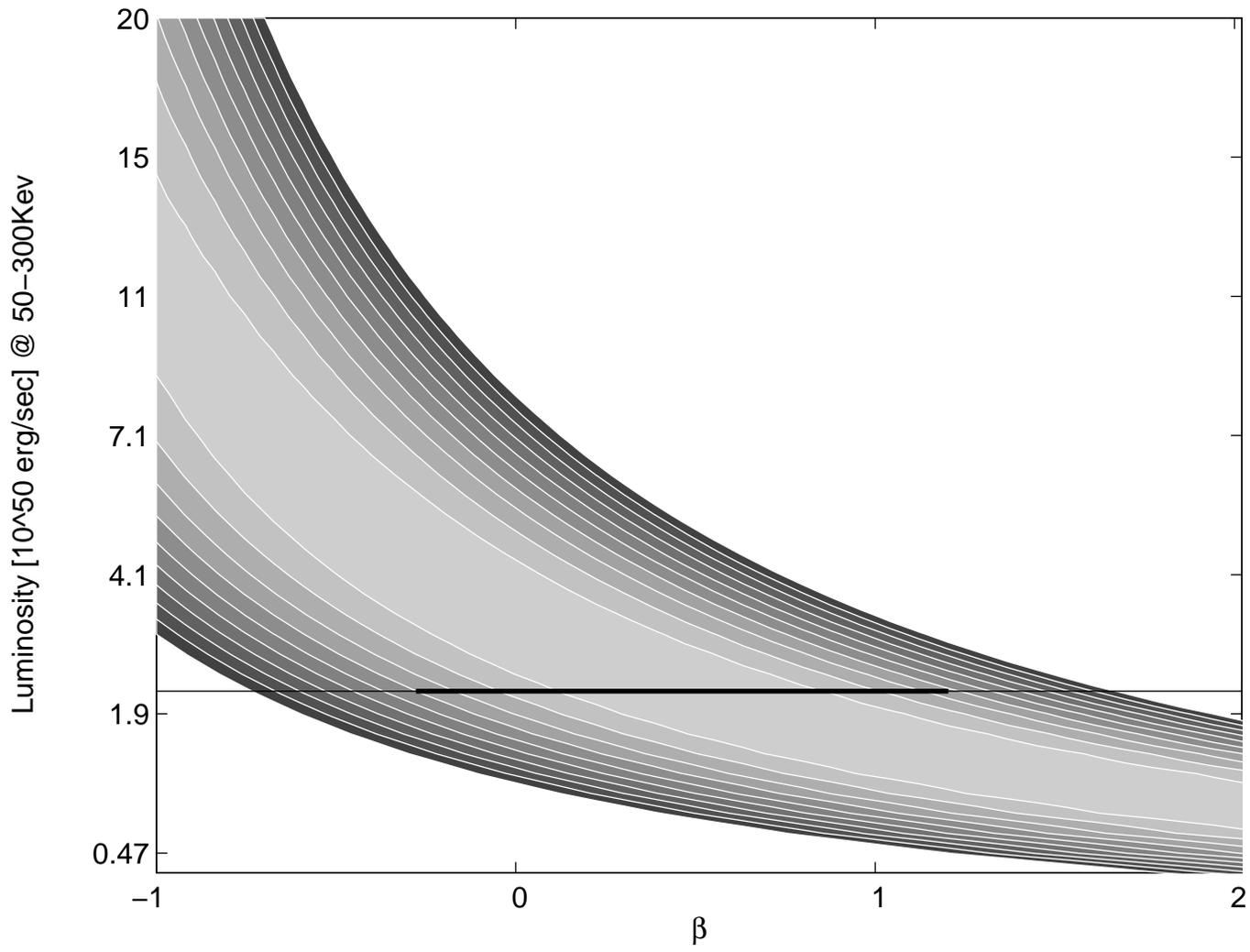
\end{document}